\documentclass[
aps,nofootinbib,
showpacs,showkeys,preprint
tightenlines,preprintnumbers,] {revtex4}

\usepackage{epsf,epsfig,subfigure,graphicx,amsmath,amssymb 
}
\usepackage{color}
\newcommand{\dis}[1]{\begin{equation}\begin{split}#1\end{split}\end{equation}}
\newcommand{\ie}{{\it i.e.~}}
\newcommand{\etal}{{\it et al.\,}}

\newcommand{\Qem}{Q_{\rm em}}

\def\sw0{{$\sin^2\theta_W^0$}}

\def\Vckm{V_{\rm CKM}}

\newcommand{\Z}{{\bf Z}}

\def\E6{{\rm E_6}}

\def\EE8{{\rm E_8\times E_8'}}

\def\flip{SU$(5)_{\rm flip}$}

\def\five{{\bf 5}}

\def\tenb{{\overline{\bf 10}}}

\def\fiveb{{\overline{\bf 5}}}

\begin{document}

\draft

\title{\Large\bf Flavor mixing inspired by flipped SU(5)  GUT}

\author{Junu Jeong$^{(1)}$, Jihn E.  Kim$^{(1,2,3)}$, Se-Jin Kim$^{(2)}$}

\address
{$^{(1)}$Center for Axion and Precision Physics Research (Institute of Basic Science), KAIST Munji Campus, 193 Munjiro,  Daejeon 34051, Republic of Korea,   \\
$^{(2)}$Department of Physics, Kyung Hee University, 26 Gyungheedaero, Dongdaemun-Gu, Seoul 02447, Republic of Korea,  \\
 $^{(3)}$ Department of Physics and Astronomy, Seoul National University, 1 Gwanakro, Gwanak-Gu, Seoul 08826, Republic of Korea 
}
 
\begin{abstract} 
We obtain a phenomenologically acceptable Cabibbo-Kobayashi-Maskawa matrix in a flipped SU(5) model inspired by the  compactification of heterotic string $\EE8$.   
 
\end{abstract}
\maketitle


\section{Introduction}\label{sec:Introduction}
``How is the current allocation of flavors realized?'' is the most urgent and also interesting one in the theoretical problems of the standard model (SM) of particle physics. Advocates of string theory for the heterotic string argue that string compactification is the most complete answer to this problem \cite{Candelas, Dixon2, Ibanez1,Tye87,Bachas87,Gepner87}.  Along this line, we study a phenomenological Cabibbo-Kobayashi-Maskawa (CKM)  mixing matrix \cite{Cabibbo63,KM73} from the recent  R-parity model  \cite{Kim18Rp} obtained from \cite{Huh09}. Other phenenological aspects are included in \cite{Kim18Rp}.

String compactifications aim at obtaining (i) large 3D space, (ii) standard-like models with three families, and (iii) no exotics at low energy (or vectorlike representations if they exist). Regarding a solution to item (i), the string landscape scenario  is suggested \cite{LandscapeSuss}, predicting about $10^{500}$ vacua for a reasonable cosmological constant  (CC).  Regarding   item (ii),   the standard-like models from heterotic string  has been suggested from early days \cite{IKNQ,Munoz88} until recently \cite{Lykken96, PokorskiW99, Cleaver99,Cleaver01, CleaverNPB,Donagi02, Raby05, He05,Donagi05,He06, Donagi06, Blumenhagen06,Cvetic06, KimJH07,Faraggi07, Blumenhagen07, Cleaver07, Munoz07,Nilles08}. Model constructions are discussed in detail in \cite{LNP696,IbanezBk,RabyBk}.
  It has been suggested that by exploring the entire string landscape one might obtain statistical data which could lead to probabilistic experimental statements \cite{Schellekens04,Anastasopoulos}. Yet the clearest statement to date is that standard-like models are exceedingly rare \cite{Lust05,LandDouglas}. In addition, the flavor problem asks for a detail model producing the observed CKM and Pontecorvo-Maki-Nakagawa-Sakada (PMNS) \cite{PMNS1,PMNS2} matrices. In this paper,  we study the flavor problem analytically in the simplest orbifold compactification based on $\Z_{12-I}$.\footnote{Among the nine orbifolds of \cite{Dixon2}, we consider $\Z_{12-I}$ is the simplest one in the sense that it has only three fixed points.} Since the number of fields are over hundred in these standard-like models,   we simplify further by choosing GUT models to ease the analytical study.   Therefore, in addition we require supersymmetry (SUSY) and  simple   or semi-simple group   grand unification (GUT) \cite{GG74,SO10GUT1,SO10GUT2,RamondE6, PS73,Barr82,DKN84}.  
  SUSY  models have been widely used to introduce a mechanism for generating a hierarchically small electroweak (EW) scale compared to the GUT scale. Above the EW scale, SUSY must be broken since no superpartner has been observed up to a TeV scale \cite{LHCichep}. In the model, therefore, SUSY breaking mechanism must be present.  The gauge group at the GUT scale is taken as $G_{\rm GUT}\times G_{\rm cond}$ where the most probable $ G_{\rm cond}$ is SU(4)$'$ \cite{NillesSU4}. In SUSY models, R-parity $P_{\mathrm {R} }=(-1)^{3(B-L)+2S}$  dictates proton stability, where $B$ is baryon number,  $L$ is lepton number, and $S$ is spin. For a conserved R-parity, it is usually assigned to a subgroup of $B-L$. From string compactification, R-parity was calculated before in this framework \cite{KimKyae07R,KimJH07,Kappl09}. Because of dangerous dimension-5 operators, leading to proton decay, $\Z_{4R}$ has been proposed in contrast to $\Z_{2R}$  \cite{Maru01,Babu03,Lee11,Tamvakis12,KimZnR}. In this paper, we work for the model of  \cite{Kim18Rp} which introduced  $\Z_{4R}$  from a string GUT.

GUTs from string compactification favor the flipped SU(5) semi-simple GUTs  \cite{Ellis89,KimKyae07,Huh09} and anti-SU(7) \cite{KimSU7}. For the simple group GUTs, SU(5), SO(10), and E$_6$, we need an adjoint representation to break the GUT groups down to the SM gauge group and it is impossible to obtain adjoint representation at the level 1 \cite{LNP696}. [Note, however, an adjoint representation of SO(10) was obtained in Ref. \cite{Tye96} at the level 3.] So, for simple studies at the level 1, anti-SU($N$) GUTs are relevant for phenomenological studies.\footnote{In Ref.  \cite{KimSU7}, anti-SU($N$) GUTs are defined as those that the GUT breaking is achieved by the anti-symmetric representations. In this definition, the flipped SU(5) is `anti-SU(5)'.} The SUSY flipped SU(5) can allow a real and symmetric $\Qem=-\frac13$ quark mass matrix, and hence the CP phase in the CKM matrix can be introduced from the $\Qem=\frac23$ quark mass matrix. This observation makes it possible to obtain the form of the CKM matrix from the string GUT.

In Sec. \ref{sec:Model}, we point out key features on the mass matrices of Ref. \cite{Kim18Rp},  and in Sec.  
 \ref{sec:Mdiagonal} we  diagonalize the mass matrices suggested in  Sec.  \ref{sec:Model}.    Three real angles of the CKM matrix are determined dominantly by the diagonalization of the $\Qem=-\frac13$ quark mass matrix, and the weak CP phase is provided from the diagonalization of the $\Qem=+\frac23$ quark mass matrix.  Neutral singlets attached to have appropriate matrix elements are listed in \cite{Kim18Rp}. 
 Sec. \ref{sec:Conclusion} is a conclusion.

\section{Mass matrices  inspired by flipped SU(5)}\label{sec:Model}

In Ref. \cite{Kim18Rp} based on the flipped SU(5) model of 
\cite{Huh09}, a possible identification $\Z_{4R}$ has been achieved, forbidding dimension-5 B violating operators but allowing the electroweak scale $\mu$ term and dimension-5 L violating Weinberg operator. The $\Z_{4R}$ quantum numbers,   $Q_{4R}$, of   the SM fields and  neutral singlets ($\sigma$'s), are presented  in Ref. \cite{Kim18Rp}.
In the flipped SU(5), the  $\Qem=-\frac13$ quarks obtain masses by the coupling
\dis{
-{\cal L}_{d}^{IJ}=f_{IJ}^{(d)} \epsilon_{ijklm} \tenb_{-1}^{I,ij}\tenb_{-1}^{J,kl}\fiveb_{+2}^{m}+{\rm h.c.}, \label{eq:dYukawa}
}
where the couplings $f_{IJ}^{(d)} $ are real parameters, $I$ and  $J$ are flavor indices and  $i,j,k,l,m$ are SU(5) indices, and the subscript is the U(1)$_X$ quantum number of \flip. $\fiveb_{+2}$ is usually denoted as $H_{dL}$ whose quantum numbers in \flip\,are  given in  Ref. \cite{Kim18Rp}.   Since interchange of the first $\tenb$ and  the second $\tenb$ in Eq. (\ref{eq:dYukawa}) is possible, the $d$-type quark mass matrix is symmetric in the weak interaction basis. $f_{IJ}$ can include vacuum expectation values (VEVs) of \flip\,singlet fields presented  in Ref. \cite{Kim18Rp}, and we choose these VEVs to be real. So, the $d$-type quark mass matrix is real and symmetric which can be diagonalized by an orthogonal matrix.
 
On the other hand, the  $\Qem=+\frac23$ quarks obtain masses by the coupling
\dis{
-{\cal L}_{u}^{IJ}=f_{IJ}^{(u)}  \tenb_{-1}^{I,ij}\five_{+3,i}^{J}\five_{-2,j}+{\rm h.c.}, \label{eq:uYukawa}
}
which need not be symmetric under the exchange $I\leftrightarrow J$.
So, the couplings $f_{IJ}^{(u)} $ need not be real parameterss, and we will assume that \flip\,singlet VEVs in $f_{IJ}^{(u)} $ can provide complex couplings. $\five_{-2}$ is usually denoted as $H_{uL}$ whose quantum numbers in \flip\,are  given in   Ref. \cite{Kim18Rp}.  So, Eq.  (\ref{eq:uYukawa}) can be diagonalized by a bi-unitrary transformation.

The effective operators of neutrino masses  in the \flip\,arise from
\dis{
\five_{+3,i}^{I}\five_{+3,j}^{J}\tenb_{-1}^{H,il}\tenb_{-1}^{H,jm}\five_{-2,l}\five_{-2,m},
}
which were discussed in \cite{Kim18Rp}.  For the PMNS matrix, the observed data are not accurate enough to analyze it here at the level of the CKM matrix. 

\section{Diagonalization of mass matrices and mixing angles}\label{sec:Mdiagonal}

With the above guidelines from string compactification, let us   parametrize the mass matrices to obtain the successful mixing matrices $V_{\rm CKM}$ and $V_{\rm PMNS}$.\footnote{$V_{\rm PMNS}$ can be discussed in parallel but we postpone the study until more accurate data on the PMNS matrix elements are determined experimentally. }   If there is a permutation symmetry $S_2$,  for the exchange $1\leftrightarrow 2$ (identifying $1\to \Phi$ and $2\to \Psi$),  we consider  the symmetric $S$ and  the antisymmetric  $A$ as
\dis{
&S=\frac{1}{\sqrt2}(\Phi+\Psi),\\
&A=\frac{1}{\sqrt2}(\Phi-\Psi).
}
Products of these two singlets are
\dis{
&AA=\frac{1}{2}(\Phi\Phi+\Psi\Psi-\Phi\Psi-\Psi\Phi),\\
&SS=\frac{1}{2}(\Phi\Phi+\Psi\Psi+\Phi\Psi+\Psi\Phi),\\
&AS=\frac{1}{2}(\Phi\Phi-\Psi\Psi ),\\
&SA=\frac{1}{2}(\Phi\Phi-\Psi\Psi ).
}
The following parametrization of mass matrices for $\Qem=-\frac13$ and $+\frac23$ quarks are used,
\dis{
&\tilde{M}_b\equiv \frac{M_{(d)}^{\rm weak}}{m_b}=\begin{pmatrix}
a_1 ,& c_1,&x_d\\
c_1,& b_1,&x_s\\
x_d ,& x_s,& 1+O(\varepsilon^{3})
\end{pmatrix}, \\[0.6cm]
&\tilde{M}_t\equiv \frac{M_{(u)}^{\rm weak}}{m_t}  =\begin{pmatrix}
a_2 ,& c_2,&x_u\\
c_2',& b_2,&x_c\\
x_u' ,& x_c',& 1+O(\eta^{3})
\end{pmatrix}. \label{eq:massMatr}
} 

The CKM matrix is given by 
\dis{
\Vckm=V_u^\dagger V_d
}
where $V_u$ and $V_d$ are diagonalizing unitary matricies of L-handed $\Qem=+\frac23$ and $\Qem=-\frac13$ quark fields. The data  for the CKM matrix is \cite{PDG18}
\dis{
|V_{\mathrm{CKM}}|\simeq 
\begin{pmatrix}
 0.97446 , & 0.22452 , & 0.00365 \\[0.5em]
 0.22438  & 0.97359 , & 0.04214  \\[0.5em]
 0.00896 , & 0.04133 , & 0.999105   
\end{pmatrix} .\label{eq:CKMpdg18}
}
\dis{
|V_{\mathrm{CKM}}|^{\rm error}\simeq 
\begin{pmatrix}
   0.00010, &   0.00044, &   0.00012 \\[0.5em]
  0.00044, &  0.00011 , &   0.00076 \\[0.5em]
  0.00024 , &  0.00074, & 0.000032
\end{pmatrix} .\label{eq:CKMpdgErr}
} 
The Wolfenstein parametrization is written as
\dis{
V^{\rm W}\simeq \begin{pmatrix}
1-\frac{\lambda^2}{2} ,& \lambda,&A\lambda^3 (\rho-i\eta)\\[0.7em ]
-\lambda,& 1-\frac{\lambda^2}{2},&A\lambda^2\\[0.7em ]
A\lambda^3 (1-\rho-i\eta),&-A\lambda^2 ,&1  
\end{pmatrix} +O(\lambda^4),
}
from which we take the following signs of the CKM elements
\dis{
V^{\rm sign}=\begin{pmatrix}
+&+&+\\ -&+&+\\ +&-&+
\end{pmatrix}.\label{eq:CKMsign}
}
A unitary matrix close to Eqs. (\ref{eq:CKMpdg18}) and (\ref{eq:CKMpdgErr}), consistent with (\ref{eq:CKMsign}), is
\dis{
V_{\rm CKM}^{\rm try}= \begin{pmatrix}
 0.974395 +8.6794\times 10^{-5} i , &  0.22481  +5.66\times 10^{-6} i , &1.41\times 10^{-3}-3.33\times 10^{-3} i , \\[0.5em]
 -0.224672-1.416\times 10^{-4} i  ,   & 0.97352- 7.46\times 10^{-5} i  , &4.23\times 10^{-2}+5.32\times 10^{-6} i \\[0.5em]
8.132\times 10^{-3} - 3.24\times 10^{-3} i , &-4.151\times 10^{-2} - 7.42\times 10^{-4} i  , & 0.99910 -4.502\times 10^{-5}   i
\end{pmatrix}  \label{eq:CKMtry}
}
To apply the Kim-Seo(KS) form \cite{KimSeo11} of the Jarlskog determinant, we check the reality of the determinant of the CKM matrix. Indeed, it is almost real:  Det$V_{\rm CKM}^{\rm try} =1-1.35525\times 10^{-20}\,i$. Also, it is almost unitary, \ie 
  $V_{\rm CKM}^{\rm try} V_{\rm CKM}^{\rm try\,\dagger}$ is
\dis{
 \begin{pmatrix}
1  ,&-2.535\times 10^{-9}-8.45\times 10^{-12}i ,& -3.63782\times 10^{-9}-6.7422\times 10^{-12}i\\[0.5em]
-2.535\times 10^{-9}+8.45\times 10^{-12}i,&1  ,& 4.922\times 10^{-9}-3.42\times 10^{-6}i \\[0.5em]
-3.63782\times 10^{-9}+6.7422\times 10^{-12}i,& 4.922\times 10^{-9}+3.42\times 10^{-6}i , &1 \\
 \end{pmatrix}
.}

Note that the following mass ratios
\dis{
\Qem=-\frac13\textrm{  quarks}:& ~~\frac{m_d}{m_b} \simeq 1.25\times 10^{-3}  ,~ \frac{m_s}{m_b} \simeq 2.5\times 10^{-2} , \\[0.5em]
\Qem=+\frac23\textrm{  quarks}: &~~\frac{m_u}{m_t} \simeq 1.4\times 10^{-5} , ~ \frac{m_c}{m_t} \simeq 0.7\times 10^{-2}.
}
  The mass hierarchy of  $\Qem=+\frac23$ quarks is more pronounced than that of  $\Qem=-\frac13$ quarks, and hence the mixing matrix of $\Qem=+\frac23$ quarks is closer to the identity than that of   $\Qem=-\frac13$ quarks.
Therefore,   the first approximation of the CKM matrix, $V_{\mathrm{CKM}} ^{\rm O}$, is set from the diagonalization of  $\Qem=-\frac13$ quark masses.
An orthogonal matrix close to $V_{\rm CKM}^{\rm try} $ of Eq. (\ref{eq:CKMtry})  is
\dis{
V_{\rm CKM}^{\rm O}= \begin{pmatrix}
 0.974034  , &  0.26385, &0.00293448~ \\[0.5em]
 -0.226277 ,   & 0.972973  , &0.0460661 \\[0.5em]
0.0075732, &-0.0455326 , & 0.998934
\end{pmatrix}  \label{eq:CKMO}
}
which gives
 Det$V_{\rm CKM}^{\rm O} =1$ and  
\dis{
V_{\rm CKM}^{\rm O} V_{\rm CKM}^{\rm O\,\dagger}=
 \begin{pmatrix}
1  ,&7.343\times 10^{-7}  ,& -7.494\times 10^{-9} \\
7.343\times 10^{-7} ,&1  ,& 1.355\times 10^{-6}  \\
-7.494\times 10^{-9},& 1.355\times 10^{-6}  , &1 \\
 \end{pmatrix}
.}
 
  For the $\Qem=-\frac13$ quark fields, the mass eigenstate basis is related to the  weak  eigenstate basis by
\dis{
q_{d\,L}^{\rm weak}= V_d q_{d\,L}^{\rm mass},~q_{d\,R}^{\rm weak}= U_d q_{d\,R}^{\rm mass}.
}  
Inspired from \flip, let  $V_d$ parametrize  (approximately) the real angles and $V_u$  determine the CP phase in the CKM matrix. Along this strategy, using the real matrix $V_{\rm CKM}^{\rm O}$ of (\ref{eq:CKMO}), we determine 
\begin{equation}
\tilde{M}_{d}\approx
V_{\mathrm{CKM}} ^{\rm O}
\begin{pmatrix}
 \frac{m_d}{m_b}, & 0, & 0 \\[0.3em]
0, &  \frac{m_s}{m_b}, & 0 \\[0.3em]
0, & 0, & 1 \\
\end{pmatrix} 
(V_{\mathrm{CKM}} ^{\rm O})^{-1}
\simeq   
\left(
\begin{array}{ccc}
 2.47579\times 10^{-3}, & 5.36634\times 10^{-3}, & 2.68287\times 10^{-3}
  \\[0.5em]
 5.36632\times 10^{-3}, & 2.58529\times 10^{-2} ,&  4.49073 \times 10^{-2}
  \\[0.5em]
2.68285\times 10^{-3}, & 4.4906\times 10^{-2} ,& 1-2.07873\times 10^{-3}\\
\end{array}
\right).\label{eq:MdSet}
\end{equation} 

 Small parameters $\varepsilon$ (for    $\Qem=-\frac13$ quarks) and $\eta$ (for    $\Qem=+\frac23$ quarks) are introduced  for the  matrices in Eq. (\ref{eq:massMatr}). For $ \varepsilon $  from the ratio of (1,3) and (2,3) elements of Eq. (\ref{eq:MdSet}), let us parametrize $\tilde{M}_{d}$ in terms of $ \varepsilon $ as 
\begin{equation}
\tilde{M}_{d}\propto 
\left(
\begin{array}{ccc}
 \alpha _{1,1} \varepsilon ^3, & \alpha _{1,2} \varepsilon^3, &  \alpha _{1,3}\varepsilon^3 \\
\alpha _{2,1} \varepsilon ^3,  & \alpha _{2,2} \varepsilon^{2} ,& \alpha _{2,3}\varepsilon^{2}  \\
\alpha _{3,1}  \varepsilon^3, &  \alpha _{3,2} \varepsilon^{2} , & 1 + \alpha _{3,3}\varepsilon ^3 \\
\end{array}\label{eq:Mdform}
\right).
\end{equation}
We chose the hierarchy $\tilde{M}_{d}(1,3)<\tilde{M}_{d}(2,3)$  and $\tilde{M}_{d}(2,1)<\tilde{M}_{d}(3,1)$ due to the symmetry properties of $A$ and $S$. 

To apply to the flipped SU(5) model discussed in  Sec. \ref{sec:Model}, we  use the following symmetrized  $\alpha_{i,j}$,\footnote{Since 
$\frac{m_s}{m_b}\approx\frac{0.1}{4}$,  we vary $\varepsilon^2$ near $\frac{1}{40}$ such that $\alpha_{2,2},\alpha_{2,3},\alpha_{3,2}$ and $\alpha_{3,3}$ turn out to be simple numbers. }  
\dis{
\begin{aligned}[c]
\textrm{For } \, \varepsilon\approx 0.160789: ~&\alpha_{1,1} \to +0.595592,\\
&\alpha_{2,1} \to +1.29096,\\
&\alpha_{3,1} \to +0.645403,
\end{aligned}
\qquad
\begin{aligned}[c]
&\alpha_{1,2} \to +1.29096,\\
&\alpha_{2,2} \to +1,\\
&\alpha_{3,2} \to +1.73698\approx \sqrt3,\\
\end{aligned}
\qquad
\begin{aligned}[c]
&\alpha_{1,3} \to +0.645408,\\
&\alpha_{2,3} \to +1.73703\approx \sqrt3,\\
&\alpha_{3,3} \to -0.500073\approx -\frac12.
\end{aligned}\\
}
We proceed to obtain a CKM matrix consistent with the above symmetric matrix. The hierarchical structure (\ref{eq:Mdform}) can be obtained as discussed in \cite{Kim18Rp}, which however is not pursued in this paper.
Then, the mass eigenvalues of $\Qem=-\frac13$ quarks are
\dis{ 
 \begin{array}{l}  e_{1} \approx  0.595592 \epsilon ^3 -1.66658\epsilon ^4-0.9926 \epsilon ^5 +O(\epsilon ^6), \\[0.5em]
 e_{2} \approx \epsilon ^2-1.3506\epsilon ^4+0.9926 \epsilon ^5 +O(\epsilon ^6), \\[0.5em]
 e_{3} \approx  1 -0.500073\epsilon ^3+3.01718\epsilon ^4 +O(\epsilon ^6),
 \end{array}  \label{eq:fit40}
}
and, in terms of   $\epsilon$ the diagonalizing matrix $V_d$ is given  by
\dis{ 
 V_d \simeq 
\left(
\begin{smallmatrix}
 \small \begin{array}{l} 1  -0.833285\epsilon ^2\\ -0.992596\epsilon ^3 -0.648831 \epsilon ^4, \end{array}
& \small \begin{array}{l} +1.29096 \epsilon +0.768885\epsilon ^2\\+
0.00471\epsilon ^3 -1.5216\epsilon ^4,  \end{array} 
& \small \begin{array}{l}   +0.64541\epsilon ^3\end{array} \\[0.5em]
 \small \begin{array}{l} -1.29096\epsilon -0.768883\epsilon ^2
 \\ -0.004693\epsilon ^3+1.52155 \epsilon ^4,\end{array}
 & \small \begin{array}{l} 1-0.83329 \epsilon ^2\\ -0.99260\epsilon ^3-2.15741 \epsilon ^4,\end{array} &
  \small \begin{array}{l} +1.73703 \epsilon ^2\\ +1.73703\epsilon ^4\end{array} \\[0.5em]
\small \begin{array}{l}  +1.59696\epsilon ^3+1.33553\epsilon ^4,\end{array} & \small \begin{array}{l} -1.73698\epsilon ^2 -1.12276\epsilon ^4, \end{array}& \small \begin{array}{l} 1  -1.50863\epsilon ^4 \end{array}\\
\end{smallmatrix}
\right).\label{eq:Vd40}
}

Similarly,   the $\Qem=+\frac23$ quark eigenstate bases can be related. However, it is more complicated than the $\Qem=-\frac13$ quark mass matrix in two aspects. Firstly, the $\Qem=\frac23$ quark mass matrix $M_u$ need not be symmetric and hence we need two unitary matrices, the L untary matrix $V_u$ and the R unitary matrix $U_u$, and second the phase $e^{i\delta}$ is introduced through $V_u$. For the diagonalization, we need $U_u$ which does not appear in the CKM matrix. For the CKM matrix, therefore, an explicit form of $U_u$ is not needed in this paper. To place the CP phase, we study phenomenologically the form of  $M_u$  when  expaned in terms of the small parametr $\eta$. To minimize the effect of the phases, let us consider a hermitian matrix
\dis{
\tilde{M}_u \tilde{M}_u^\dagger= V_u \begin{pmatrix}
({m_u}/{m_t})^2,&0,&0~\\[0.5em]
0,&{m_c}/{m_t})^2, &0~\\[0.5em]
 0,&0,&1~\\[0.5em]
\end{pmatrix} V_u^\dagger .\label{eq:Mu2}
}
Because the R-unitary matrix does not appear explicitly in the CKM matrix, we employ the freedom on $U_u$. Namely, we use the hierarchy for the hermitian matrix $\tilde{M}_u \tilde{M}_u^\dagger$, but $\tilde{M}_u$ may not have the hierarchical form, as we will see later. Namely, we use the freedom of $\tilde M_u$ in obtaining the observed $\Vckm$.
Using
\dis{
V_u=V_d \Vckm^\dagger,\label{eq:Vu}
}
with $V_d$ of (\ref{eq:Vd40}) for $\varepsilon=0.16$,  $V_u$ can be written as
\dis{
\tilde{M}_u \tilde{M}_u^\dagger=\begin{pmatrix}
1.31795\cdot 10^{-5},& 1.38244\cdot 10^{-5}\cdot e^{i(\frac{\pi}{2}-0.4153)}, &3.63026\cdot 10^{-3}\cdot e^{i(\frac{\pi}{2}-0.4153)}\\[0.5em]
1.38244\cdot 10^{-5}\cdot e^{i(-\frac{\pi}{2}+0.4153)},&6.34168\cdot 10^{-5} ,&3.79682\cdot 10^{-3}\cdot e^{i(-0.00248)}\\[0.5em]
3.63026\cdot 10^{-3}\cdot e^{i(-\frac{\pi}{2}+0.4125)},& 3.79682\cdot 10^{-3}\cdot e^{i(0.00248)},&1-2.7465\cdot 10^{-5}
\end{pmatrix}\label{eq:VuNum}
}
from which we can calculate the matrix multiplication of two $(\tilde{M}_u \tilde{M}_u^\dagger)$'s
\dis{
&(\tilde{M}_u \tilde{M}_u^\dagger)_{1,1}-(\tilde{M}_u \tilde{M}_u^\dagger)_{1,3}(\tilde{M}_u \tilde{M}_u^\dagger)_{1,3}^*= 6.97823\cdot 10^{-10},\\[0.3em]
&(\tilde{M}_u \tilde{M}_u^\dagger)_{1,2}-(\tilde{M}_u \tilde{M}_u^\dagger)_{1,3}(\tilde{M}_u \tilde{M}_u^\dagger)_{2,3}^*= 8.25259\cdot 10^{-8}-8.548\cdot 10^{-8}i,\\[0.3em]
&(\tilde{M}_u \tilde{M}_u^\dagger)_{2,2}-(\tilde{M}_u \tilde{M}_u^\dagger)_{2,3}(\tilde{M}_u \tilde{M}_u^\dagger)_{2,3}^*= 4.9\cdot 10^{-5}.
}
Thus, we obtain the following approximate relations
\dis{
&\tilde{M}_{u}(1,3)\approx (\tilde{M}_u \tilde{M}_u^\dagger)_{1,3}\,(\textrm{from the assumption } \tilde{M}_{u}(3,3)\approx 1),\\
&\tilde{M}_{u}(2,3)\approx (\tilde{M}_u \tilde{M}_u^\dagger)_{2,3},\\
& \tilde{M}_{u}(2,1)\sim  \tilde{M}_{u}(2,2)\sim  \tilde{M}_{u}(2,3).\label{eq:MuApprox}
}
 Parametrizing $\tilde{M}_{u}$ in terms of $ \eta $ as 
\begin{equation}
\tilde{M}_{u}\propto 
\left(
\begin{array}{ccc}
\eta^4 \beta_{1,1} , & \eta^4 \beta_{1,2}e^{i\delta_{1,2}} , & \eta^2 \beta _{1,3}e^{i\delta_{1,3}} \\[0.3em]
\eta^2 \beta_{2,1}e^{i\delta_{2,1}},  &\eta^2 \beta_{2,2},&\eta^2 \beta_{2,3}e^{i\delta_{2,3}}  \\[0.3em]
\eta^4 \beta_{3,1}e^{i\delta_{3,1}}, &  \eta^4 \beta_{3,2}e^{i\delta_{3,2}} , & 1 + \eta^4 \beta_{3,3} \\
\end{array} 
\right),~\tilde{M}_{u}^\dagger\propto 
\left(
\begin{array}{ccc}
\eta^4 \beta_{1,1} , & \eta^2 \beta_{2,1}e^{-i\delta_{2,1}} , & \eta^4 \beta _{3,1}e^{-i\delta_{3,1}} \\[0.3em]
\eta^4 \beta_{1,2}e^{-i\delta_{1,2}},  &\eta^2 \beta_{2,2},&\eta^4 \beta_{3,2}e^{
-i\delta_{3,2}}  \\[0.3em]
\eta^2 \beta_{1,3}e^{-i\delta_{1,3}}, &  \eta^2 \beta_{2,3}e^{-i\delta_{2,3}} , & 1 + \eta^4 \beta_{3,3} \\
\end{array}\right),
\label{eq:Muform}
\end{equation}
we obtain
\dis{
\tilde{M}_u \tilde{M}_u^\dagger=
\begin{pmatrix}
\eta^4 \beta_{1,3}^2 , & \eta^4 \beta_{1,3}\beta_{2,3}e^{i(\delta_{1,3}-\delta_{2,3} )} , & \eta^2 \beta _{1,3}e^{i\delta_{1,3}} \\[0.5em]
\eta^4 \beta_{1,3}\beta_{2,3}e^{-i(\delta_{1,3}-\delta_{2,3} )},  &\eta^4[ (\beta_{2,1})^2+(\beta_{2,2})^2],&\eta^2 \beta_{2,3}e^{i\delta_{2,3}}  \\[0.5em]
\eta^2 \beta_{1,3}e^{-i\delta_{1,3}}, &  \eta^2 \beta_{2,3}e^{-i\delta_{2,3}} , & 1  \\
\end{pmatrix} \label{eq:MuMu}
}
From Eqs.   (\ref{eq:MuApprox}), (\ref{eq:Muform}) and (\ref{eq:MuMu}), we obtain
\dis{
\tilde{M}_u\simeq \begin{pmatrix}
\eta^4 \beta_{1,1} , & \eta^4 \beta_{1,2}e^{i\delta_{1,2}} , &3.63026\cdot 10^{-3}\cdot e^{i(\frac{\pi}{2}-0.4153)} \\[0.5em]
\eta^2 \beta_{2,1}e^{i\delta_{2,1}},  &\eta^2 \beta_{2,2},&3.79682\cdot 10^{-3}\cdot e^{i(-0.00248)} \\[0.5em]
\eta^4 \beta_{3,1}e^{i\delta_{3,1}}, &  \eta^4 \beta_{3,2}e^{i\delta_{3,2}} , & 1 + \eta^4 \beta_{3,3} \\
\end{pmatrix}\label{eq:MuNumer}
}
with $\eta^2\sqrt{(\beta_{2,1})^2+ (\beta_{2,2})^2}=0.700\times 10^{-2}$ such that  $ \beta_{2,1}^2+\beta_{2,2}^2\approx 1$ and $\eta\approx   0.0837$. Thus, we obtain
\dis{
& \beta_{1,3}\simeq 0.518604,\\
& \beta_{2,3}\simeq 0.542398.
}

Now, let us obtain $\tilde{M}_u$   from 
\begin{equation}
\tilde{M}_{u}\approx
V_u
\begin{pmatrix}
 \frac{m_u}{m_t}, & 0, & 0 \\[0.3em]
0, &  \frac{m_c}{m_t}, & 0 \\[0.3em]
0, & 0, & 1 \\
\end{pmatrix} 
U_u^\dagger ,\label{eq:MuSet}
\end{equation} 
where $V_u$ is given in Eq. (\ref{eq:VuNum}) and $U_u$ is
\dis{
\begin{pmatrix}
  R_{1,1},& R_{1,2},& R_{1,3}\\
 R_{2,1},& R_{2,2},& R_{2,3}\\
 R_{3,1},& R_{3,2},& R_{3,3}\\
\end{pmatrix}.
}
Thus, $\tilde{M}_{u}$ becomes
\dis{
&\begin{pmatrix}
 1.40\times 10^{-5},& 1.178\times 10^{-5}\cdot e^{i(0.09211)},&  3.6303\times 10^{-3}\cdot e^{i(\frac{\pi}{2}- 0.4125)}\\
 2.36\times 10^{-8}\cdot e^{i(\pi-0.099)}, & 7.000\times 10^{-3}\cdot e^{i(0.0000)},  & 3.80\times 10^{-3}\cdot e^{i(-0.0024)}, \\
 5.07810\times 10^{-8}\cdot e^{i(\frac{\pi}{2}+0.4109)} ,&  2.66\times 10^{-5}\cdot e^{i(\pi+0.0020)},  & 0.99999\\
\end{pmatrix} U_u^\dagger \\[0.7em]
&~~~=\left(u_1,u_2,u_3 \right)\label{eq:Ui}
}
where $u_i$ is
\dis{
\begin{array}{l}
 [3.6303\cdot e^{i(\frac{\pi}{2}- 0.4125)} (R_{i,3})^* ]\times 10^{-3}\\[0.5em]
 [ 7.00(R_{i,2})^* +3.80 \cdot e^{i(-0.0024)}(R_{i,3})^*] \times 10^{-3} \\[0.5em]
 0.9999 (R_{i,3})^*+2.66\cdot e^{i(\pi+0.0020)} (R_{i,2})^* \times 10^{-5}.
 \end{array}\label{eq:URPhase}
}
Existence of $U_u$ is sufficient for our study of the CKM matrix.  In this regard, note that an R-hand unitary matrix $U_u$ close to
\dis{
U_u\simeq \begin{pmatrix}
 \frac{1}{2},&- \frac{\sqrt3}{2},& 0\\[0.5em]
\frac{\sqrt3}{2},& \frac{1}{2},& 0\\[0.5em]
0,& 0,&1\\
\end{pmatrix}\label{eq:RRreal}
}  
gives a solution  (\ref{eq:Ui}) consistent with the data  (\ref{eq:MuNumer}). Later, we will obtain the CP phase of the CKM matrix is close to $\frac{\pi}{2}$. Let us suppose that $U_u$ has 0 entries as shown in (\ref{eq:RRreal}).  The first line of Eq. (\ref{eq:URPhase}) is useful since there is only one term and a statement on the phase is clear. From (\ref{eq:URPhase}), then if $ (R_{3,3})^* $ has a phase close to $\frac{ \pi}{8}\approx 0.393$, then the   element $\tilde{M}_u(1,3)$ has a phase $\frac{\pi}{2}-0.02$. So, the $\Qem=\frac23$ quark mass matrix can have a phase close to the observed CP phase of the CKM matrix, which can help constructing a field theoretic model.

What we obtain from  $V_d$ of (\ref{eq:Vd40}) and  $V_u$ given in (\ref{eq:VuNum}) is the same as Eq. (\ref{eq:CKMtry}),
\dis{
V_{\rm CKM}=V_u^\dagger V_d =  \begin{pmatrix}
+0.974395 \cdot e  ^{i(8.90745\times 10^{-5}) }  ,&+0.224814 \cdot e  ^{i(2.51923\times 10^{-5}) }  ,& +0.003615 \cdot e  ^{i(-\frac{\pi}{2} +0.4005) } \\[0.5em]
-0.224672 \cdot e  ^{i(6.302\times 10^{-4}) } ,&+0.973517 \cdot e  ^{i(-7.666\times 10^{-5}) } ,&+0.042275 \cdot e  ^{i(1.258\times 10^{-4}) }   \\[0.5em]
+0.008754 \cdot e  ^{i(-0.37945) } ,&-0.041516 \cdot e  ^{i(1.788\times 10^{-2}) } ,&0.99910 \cdot e  ^{i(-4.506\times 10^{-5}) } \\
 \end{pmatrix}.
}
Then, the KS form \cite{KimSeo11}  for the Jarlskog determinant is 
\dis{
J=|{\rm Im}\,V_{31} V_{22} V_{13}| &=|(0.008754)\cdot (0.973517)\cdot (0.003615)\cdot
\sin (-0.37945-7.666\times 10^{-5} -\frac{\pi}{2} +0.4005)|\\
&\simeq |3.081\times 10^{-5}\cdot \sin(-88.8^{\rm o})|\simeq 3.08\times 10^{-5}.
}
which is consistent with the value of the Particle Data Group, $J_{PDG}=(3.18\pm0.15)\times 10^{-5}$ \cite{PDG18}. Here, $\alpha$ of the unitarity triangle is $88.8^{\rm o}$.
 
\section{Conclusion}\label{sec:Conclusion} 

Starting from a \flip\,inspired mass matrices \cite{Kim18Rp}, we obtained the  CKM matrix which can be made consistent with the observed data \cite{PDG18}.  The model presented in Ref. \cite{Kim18Rp} allows a $\Z_{4R}$ discrete symmetry such that it forbids the dimension-5 B violating operators but allows the needed electroweak scale $\mu$ term and dimension-5 lepton number violating Weinberg operators.

\acknowledgments{ This work is supported in part by the IBS (IBS-R017-D1-2014-a00) and by the National Research Foundation (NRF) grant  NRF-2018R1A2A3074631, and in addition S-J.K. is supported in part by NRF-2018R1D1A1B07045414.}


\end{document}